\documentclass[article,12pt]{elsarticle}
\usepackage[titletoc]{appendix}
\usepackage{ucs}
\usepackage[utf8x]{inputenc}
\usepackage[scaled=1]{couriers}
\usepackage[T1]{fontenc}
\usepackage[english,spanish]{babel}
\usepackage{amsmath, amsthm, amssymb}
\usepackage{textcomp}
\usepackage{color}
\usepackage{calc}
\usepackage{longtable}
\usepackage{lscape}
\usepackage{supertabular}
\usepackage{gensymb}
\usepackage{MnSymbol} 
\usepackage{lscape}
\usepackage{wasysym}
\usepackage{hyperref}
\usepackage{tabulary}
\usepackage{array}
\usepackage{hhline}
\usepackage{ulem}
\usepackage{lineno}
\usepackage{graphics}
\usepackage{natbib}
\usepackage{hypernat}

\newtheorem{theorem}{Theorem}

\hypersetup{colorlinks=true, linkcolor=blue, filecolor=blue, urlcolor=blue}

\makeatletter

\newcommand\ps@Standard{%
\renewcommand\@oddhead{}%
\renewcommand\@evenhead{}%
\renewcommand\@oddfoot{\thepage}%
\renewcommand\@evenfoot{\@oddfoot}%
\setlength\paperwidth{8.5in}
\setlength\paperheight{11in}
\setlength\voffset{-1in}
\setlength\hoffset{-1in}
\setlength\topmargin{0.7874in}
\setlength\headheight{12pt}
\setlength\headsep{0cm}
\setlength\footskip{12pt+0.1965in}
\setlength\textheight{11in-0.7874in-0.7874in-0cm-12pt-0.1965in-12pt}
\setlength\oddsidemargin{0.7874in}
\setlength\textwidth{8.5in-0.7874in-0.7874in}
\renewcommand\thepage{\arabic{page}}
\setlength{\skip\footins}{0.0398in}
\renewcommand\footnoterule{\vspace*{-0.0071in}
\noindent\textcolor{black}{\rule{0.25\columnwidth}{0.0071in}}
\vspace*{0.0398in}}
}
\makeatother
\journal{Progress in Biophysics and Molecular Biology}

\begin{document}
\selectlanguage{english}



\citestyle{numbers} 
\citestyle{square} 
\citestyle{citesep={,}} 
\citestyle{sort}
\citestyle{compress} 
\bibliographystyle{unsrtnat}   

\begin{frontmatter}
\title{Caveat on the Boltzmann distribution function use in biology.}

\author{Carlos Sevcik\corref{cor}}

\address{Laboratory on Cellular Neuropharmacology, Centro de Biof\'isica y Bioqu\'imica, Instituto Venezolano de Investigaciones Cient\'ificas (IVIC), Caracas, Venezuela.}

\cortext[cor]{Prof. Carlos Sevcik, IVIC CBB, Apartado 20632, Caracas 1020A, Venezuela. Email: csevcik@ivic.gob.ve. Private Email: carlos.sevcik.s@gmail.com. Phone: +58 212 504 1399, Fax: +58 212 504 1093. Mobile: +58 412 931 9162}
\begin{abstract}
Sigmoid semilogarithmic functions with shape of Boltzmann equations, have become extremely popular to describe diverse biological situations. Part of the popularity is due to the easy availability of software which fits Boltzmann functions to data, without much knowledge of the fitting procedure or the statistical properties of the parameters derived from the procedure. The purpose of this paper is to explore the plasticity of the Boltzmann function to fit data, some aspects of the optimization procedure to fit the function to data and how to use this plastic function to differentiate the effect of treatment on data and to attest the statistical significance of treatment effect on the data. 
\end{abstract}


\end{frontmatter}

\section{Introduction.}

\begin{footnotesize}
\begin{tabular}{*{2}{m{8cm}}}
\textquotedblleft{}When a finger points at the moon, \par one must not mistake the finger with the moon.\textquotedblright \par \text{    } \par Old Zen Buddhist advice.
&
\textquotedblleft{}With four parameters I can fit an elephant \par  and with five I can make him wiggle his trunk.\textquotedblright\par \text{     } \par Attributed to J. Von Neuman \citep{Dyson2004}.
\\
\end{tabular}
\end{footnotesize}
\\

Some thoughts and some equations transcending their greatness, also transcend their purpose. During the last quarter of the XIX century Ludwig Boltzmann \citep[Ch. 9]{Moore1972} derived an equation which predicts the proportion $ N_j $ particles, in an ensemble of $ N $ non interacting particles, that are in a state with particle energy $ \epsilon_j $
\begin{equation}\label{E:Boltzmann}
\frac{N_j}{N}=\frac{e^{-\epsilon_j/kT}}{\sum_{j=1}^{N_j}e^{-\epsilon_j/kT}} = \frac{e^{-\epsilon_j/kT}}{\zeta(T)}
\end{equation}
where $ \zeta (T)$ is called the\textit{ particle partition function} or when dealing with molecules, the \textit{molecular partition function}. When only two energy levels are dealt with, the ratio of $ N_0 $ particles in energy level $ \epsilon_0 $ with $ N_1 $ particles in energy level $ \epsilon_1 $ predicted by equation \ref{E:Boltzmann} is
\begin{equation}\label{E:BolzTwoStates}
\frac{N_1}{N_0}= e^{-(\epsilon_1-\epsilon_0)/kT}.
\end{equation}

In an extension of eq, (\ref{E:Boltzmann}) for \textit{degenerate} systems (when more then one states $ \epsilon_j$  have the same energy) a statistical weight, $ g_j $ equal to the number of superimposed levels, is included. Then
\begin{equation}\label{E:BoltzDegen}
\frac{N_j}{N}=\frac{g_j e^{-\epsilon_j/kT}}{\sum_{j=1}^{N_j}g_j e^{-\epsilon_j/kT}}.
\end{equation}
Which is the Boltzmann distribution law in its most general form. The average kinetic energy is
\begin{equation}\label{E:MeanEnerg}
\bar{\epsilon}=\frac{\sum_{j=1}^{N_j}N_j  \epsilon_j}{\sum_{j=1}^{N_j} N_j} = \frac{\sum_{j=1}^{N_j}\epsilon_j g_j  e^{-\epsilon_j/kT}}{\sum_{j=1}^{N_j} g_j e^{-\epsilon_j/kT}}=kT^2 \left (\frac{\partial \ln \zeta}{\partial T}\right)_V.
\end{equation}

The middle term in Eq. (\ref{E:MeanEnerg}) includes the \textit{statistical weights} $g_i$ which account for, so called, \textit{degenerate} levels. The molecular partition function is useful only when the system of interest can be considered to be made up of noninteracting particles, molecules with no appreciable intermolecular forces. Only then, can we define and enumerate the states of the system in either terms of quantum mechanical energy states of individual molecules, or classical positions and moments of individual molecules. When interactions between molecules occur, the description of the states of the system must include potential energy terms, such as  $ U(r_{ij}) $, which are functions of intermolecular distances.

Equation (\ref{E:Boltzmann})  may be rewritten for ensembles of interacting particles as
\begin{equation}\label{E:MaxwBoltz}
p_j=\frac{N_j}{N}=\frac{e^{-E_j/kT}}{\sum_{j=1}^{N_j}e^{-E_j/kT}} = \frac{e^{-E_j/kT}}{Z(T)}
\end{equation}
where $ E_j = \frac{1}{2} m v^2 + U_j $, $m$ is the particle mass, $v$ is velocity and $U_j$ is potential energy. Equation (\ref{E:MaxwBoltz}) is the \textit{Maxwell-Boltzmann} distribution function. If there are only two possible states in the system Eq. (\ref{E:MaxwBoltz}) becomes
\begin{equation}\label{E:Common}
\frac{p_1}{p_1 + p_2}=\frac{e^{-E_1/kT}}{e^{-E_1/kT}+e^{-E_2/kT}}  \implies  p_1=\frac{1}{1+e^{-(E_2-E_1)/kT}}.
\end{equation}

 When dealing with cell membranes, a Boltzmann equation is expressed as free energy in voltage units (the electrical potential difference existing across cell membranes) in general the form used looks like
\begin{equation}\label{E:BoltzVolt}
B= \frac{1}{1+e^{-(V_2-V_1)/\kappa}}
\end{equation}
where $V$ is used to more clearly specify that we deal with electrical potential differences, $\kappa$ is customarily referred to as \textquotedblleft slope factor\textquotedblright{ }(see for example \citet{Peigneur2012}). Since these situations deal with ensembles of particles the{ }\textquotedblleft slope factor\textquotedblright{ }is usually parametrized $\kappa =RT/zF \approx 25.4 \text{ mV}$, at room temperature if: the ionic valence, \textit{z}=1, \textit{F} is the Faraday constant, \textit{R} is the gas constant and \textit{T} is the absolute temperature. In electrophysiology, an equation of the form (\ref{E:BoltzVolt}) was introduced for the first time by \citet{Hodgkin1952d, Hodgkin1952e} (H\&H for brevity), and was used to describe the distribution, inside or outside axons, of hypothetical particles associated with Na\textsuperscript{+} and K\textsuperscript{+} currents crossing the nerve membrane. Since the \citet{Hodgkin1952e} work was seminal for electrophysiology, a plethora of papers have used Boltzmann functions in connection with electrical properties of cells and isolated ionic channels \citep{Sakmann1984}. In electrophysiology, however equation (\ref{E:BoltzVolt}) is modified \citep{Hodgkin1952d}, $V_2$ represent \textit{cell trans membrane potential} (\textit{membrane potential} for short) expressed plainly as $V$ and since Eq. (\ref{E:BoltzVolt}) takes values between 0 and 1, $V_2$  is taken as the membrane potential where $B=0.5$ and is usually termed $ V_{1/2} $, Eq. (\ref{E:BoltzVolt}) thus becomes
\begin{equation}\label{E:BoltzElect}
B \left( V | V_{\text{\textonehalf}},\kappa \right)= \frac{1}{1+e^{-(V-V_{1/2})/\kappa}}.
\end{equation}
When Eq. (\ref{E:BoltzElect}) is used in the original fashion of H\&H, to represent trans membrane distribution of some charged particle, $B$ is expressed in respect to the potential at which 50\% of the particles are in one side of the membrane, and 50\% is at the other side. Eq. (\ref{E:BoltzElect}) is thus reduced to a situation where a dependent variable $B$ may be fitted by some nonlinear optimization procedure to an independent variable $V$ (usually expressed in mV) using Eq. (\ref{E:BoltzElect}). The optimization procedure enables to estimate the parameters $V_{1/2}$ and $\kappa$. In  H\&H work \citep[pg 501, Eq. 1]{Hodgkin1952d}, 
\begin{equation}\label{E:h_infin}
h_{\text{steady state}} = \frac{1}{1+e^{-(V-V_{h}) / 7}}
\end{equation}
which is presented here with post H\&H membrane potential sign conventions. $V_h$ was estimated to be close to the resting membrane potential. H\&H also used a Boltzmann function to estimate properties of hypothetical particles gating or triggering the mechanism controlling Na\textsuperscript{+} conductance in nerve \citep[PART I, pp. 503--504]{Hodgkin1952e} the form of the Boltzmann function in this case was
\begin{equation}\label{E:Gating1}
P_i=\frac{1}{1+e^{-(w+z \epsilon E)/kT}}
\end{equation}
where $E$ is trans membrane potential, $w$ is work done while molecule from the inside move to the outside of the membrane, $z$ is valence of the molecule or number of positive charges on the molecule and $\epsilon$ is the absolute charge of an electron. With large $E$, Eq. (\ref{E:Gating1}) was used in the limit form
\begin{equation}\label{E:GatingLim}
P_i=K e^{z \epsilon E}
\end{equation}
to estimate $z$ of hypothetical particles; in Ec (\ref{E:GatingLim}), $ K $ is a constant. Quoting \citet[PART I, pp. 503--504]{Hodgkin1952e}:
\begin{quotation}
\begin{footnotesize}
\noindent{}\textquotedblleft{}whose distribution changes must bear six negative electronic charges, or, if a similar theory is developed in terms of the orientation of a long molecule with a dipole moment, it must have at least three negative charges on one end and three positive charges on the other. A different but related approach is to suppose that sodium movement depends on the presence of six singly charged molecules at a particular site near the inside of the membrane.\textquotedblleft{}
\end{footnotesize}
\end{quotation}
but  
\begin{quotation}
\begin{footnotesize}
\noindent{}\textquotedblleft{}Details of the mechanism will probably not be settled for some time, but it seems difficult to escape the conclusion that the changes in ionic permeability depend on the movement of some component of the membrane which behaves as though it had a large charge or dipole moment. If such components exist it is necessary to suppose that their density is relatively low and that a number of sodium ions cross the membrane at a single active patch.\textquotedblright{}
\end{footnotesize}
\end{quotation}
In modern terms, $h_{\text{steady state}}$ expresses \textit{availability} of sodium channels to be activated when the membrane depolarizes. Yet, the term \textit{channel} was never used by \citet{Hodgkin1952b, Hodgkin1952c, Hodgkin1952d, Hodgkin1952e} in their famous series of papers, which never mentioned the term \textit{probability} in a statistical sense. When H\&H work was carried out, the existence of cellular plasma membranes has not been proven, and nothing was known on the nature of macromolecules associated with nerve excitation. The \citet{Hodgkin1952e} model is an outstanding example of model fitting to data based on bright intuition. It took close to 20 years for evidence to appear proving the existence of gating currents due to charge movements preceding channel activation as predicted by H\&H intuition \citep{Armstrong1974}. 

Equations of the form of Eq. (\ref{E:BoltzElect}) have become extremely popular to describe diverse biological situations \citep{Hodgkin1952d, Rouzaire1991, Gavallotti2008, Fogolari2002, Koegel2003, Sun2005, Habela2007, Schuster2008, Zhou2008, Cambien2008, Dubois2009}, where it is referred as a \textquotedblleft{}Boltzmann function\textquotedblright{}  or as a \textquotedblleft{}Boltzmann equation.\textquotedblright{ } Part of the popularity is the easy availability of software which fits Boltzmann functions to data, without much knowledge of the fitting procedure or the statistical properties of the parameters derived from the procedure. The purpose of this paper is to explore the plasticity of the Boltzmann function to fit data, some aspects of the optimization procedure to fit the function to data and on how to use this plastic function to differentiate the effect of a \textquoteleft{}treatment\textquoteright{} (anything that may change the system) on data, when effect is marred by uncertainty. The stress in this review is to point out is the (Proper?) use of the Boltzmann function in situations where it is related to the mechanism underlying the process under study (a few examples are considered) from situations where is given a mechanistic role based in just curve fitting. The review also pretends to stress the fact that when equation (\ref{E:Common}) is fitted to experimental data, it is frequently forgotten that this implies a nonlinear fit. When determining parameters that are not linearly independent, one parameter estimate uncertainty \textquotedblleft{}seeps\textquotedblright{} into the uncertainty of the other parameter; formally: the autocovarience of the parameters is not null. To achieve this purpose the review includes some description of Boltzmann’s function statistical properties, which are seldom described and often ignored when this function is used in biology.

\section{Methods.}

\subsection{Monte Carlo Boltzmann functions simulation.}\label{S:MonteCarlo}

To test the goodness of fitting curves to data, random data with known statistical properties were generated using Monte Carlo simulation \citep{Dahlquist1974}. For this purpose sets of pairs $(V_i, b'_i(V_i))$ were generated at $\{V_i\} = \{-100,-90, \ldots, 90, 100\}$ (in the fashion of \citet[figures 4 -- 7]{Forsyth2012}) with $b'_i(V_i)$ defined using $b_i(V_i)$ like in eq. (\ref{E:ExpBoltz}), and
\begin{equation}\label{E:RanNor}
b'_i(V_i) = N(0,1) \sigma + b_i(V_i)=N(b_i(V_i),\sigma^2).
\end{equation}
Thus the random normal variate $b'_i(V_i)=N(b_i(V_i),\sigma^2)$ contains $N(0,1)\sigma = N(0,\sigma^2)$ as $\epsilon_i$ [Eq. (\ref{E:RanVar})], the \textquotedblleft{}experimental error\textquotedblright. 

The uniform pseudo-random variables in the closed interval $[0,1]$ [$U[0,1]$] required by the \citet{Box1958} algorithm to generate $N(0,1)$, were produced using the Mersenne Twister of \citet{Matsumoto1998} with improved initialization (www.math.sci.hiroshima-u.ac.jp/~m-mat/MT/MT2002/emt19937ar.html) coded in GNU C++ (g++ v. 4.9.2), the seed required to initialize the algorithm was obtained from the Ubuntu Linux 15.04 entropy gathering device \textbf{\textit{/dev/random}}. A copy of the program source is provided in a digital annex to this manuscript. For the purpose of the simulations, initially, $\kappa$ was set to 6 mV and $V_{\text{\textonehalf}}$ was set to -40 mV, the choices are arbitrary and irrelevant, they approximate with modern conventions the values for nerve determined by H\&H and may be taken as a small tribute to their monumental work after its 60\textsuperscript{th} anniversary. The values chosen may easily be replaced by any other as long as the variance modeled is kept.

\subsection{Statistical Procedures.}

Curves were adjusted to data using a simplex minimization \citep{Nelder1965}. The statistical significance analysis of the differences between Boltzmann curves was done using \citet{Kolmogorov1933} statistics as described in sections \ref{S:CompBoltz}. Results are presented as medians and their 95\% confidence interval (CI) determined with the Hodges and Lehman method. Comparisons of two samples were done with the nonparametric Mann-Whitney (Wilcoxon) test; for all nonparametric statistical procedures please refer to \citet{Hollander1973}. Statistical differences between samples were considered significant when the probability that they stem from chance was $ \leqq 5\% $ ($ P  \leqq  0.05 $). All statistical procedures were carried out using the program \textit{est } which is freely available in C++ source form, and compiled for Linux, Windows and Macintosh OS X from \text{ftp://toxico.ivic.gob.ve/estadistica/Stat\_package}.

\section{Results.}

\subsection{The plasticity of the Boltzmann proportion function to fit data.}\label{S:Plasticity}
\begin{figure}
\centering
\includegraphics[width=12cm]{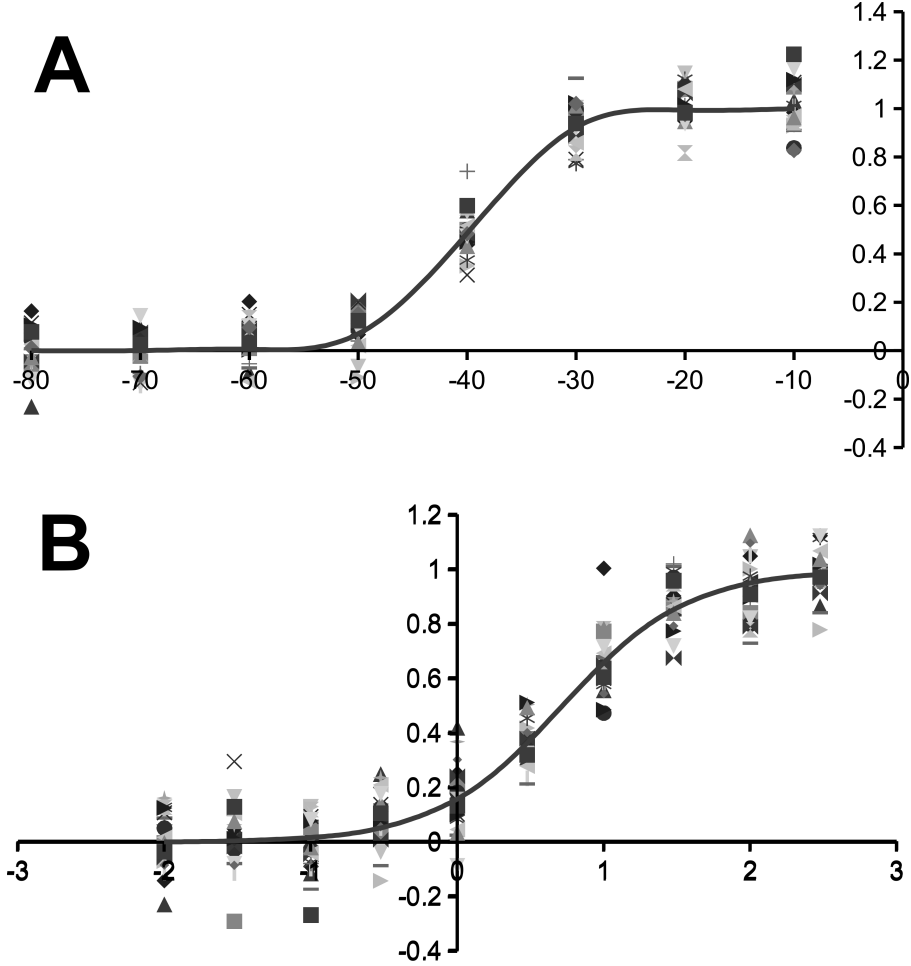}
\caption{\textbf{The plasticity of the Boltzmann function to fit data.} Symbols in \textbf{panel A }represent data produced as $y_i = 1 / (1-\exp[-(x-V_{1/2})/\kappa] + N(0,0.1)$ calculated, setting $V_{1/2} = -40$ and $\kappa = 4$, for 20 data sets with abscissa data points $\{x_i\}=\lbrace -100, -90, -80, \dots, -10 \rbrace$.  Symbols in \textbf{panel B} represent data produced as $y_i = 1 / (1+k/x) + N(0,0.1)$  calculated with $ k= 5$, for 20 data sets with abscissa data points $\{x_i\}=\lbrace 0.01, 0.03, 1, 3, 10, 30 , 100 , 300 \rbrace$ plotted vs $\log_{10}(x)$, plotted \textit{vs} x.  In each case 20 sets of 10 data pairs were produced by Monte Carlo simulation and wer e fitted to Eq. (\ref{E:BoltzElect}) with a simplex algorithm \citep{Nelder1965}. Curves fitted are presented as solid lines drawn through the data points. Parameters calculated with the simplex algorithm describing the curves were: \textbf{Panel A}, $V_{1/2} = -39.7576$ and $\kappa = 3.95265$, calculated for $n = 200$ pairs; \textbf{Panel B}, $V_{1/2} = 0.716864$ and $\kappa = 0.430228$, calculated for $n = 200$ pairs. Other details in the text of the communication.}
\label{fig:Figure1}
\end{figure}

To show the plasticity of the Boltzmann function to fit data, 2 groups of 20 sets of data were generated using Monte Carlo simulation as explained before (Section \ref{S:MonteCarlo}). The first group  was simulated about a Boltzmann function [Eq. (\ref{E:BoltzElect})] with uncertainty as follows
\begin{equation}
y_i= \frac{1}{1+e^{-(x_i-V_{1/2})/\kappa}} + N(0,0.1) 
\end{equation}
for 20 data sets with abscissa data points $\{x_i\}=\lbrace -100, -90, -80, \dots, -10 \rbrace$, with $V_{1/2} = -40$ and $\kappa = 4$; data generated are presented as diverse symbols in Fig. \ref{fig:Figure1}A. The solid line in Fig. \ref{fig:Figure1}A obeys Eq.  (\ref{E:BoltzElect}), with $V_{1/2}=-39.7576$ and $\kappa=3.95265$, both values determined with the simplex algorithm for $n = 200$ pairs of simulated data.

As an example of arbitrary use of the Boltzmann curve to fit unrelated data, the second group was simulated about a rectangular hyperbola \citep{Hill1910b,  Adair1925b, Langmuir1918} 
\begin{equation}\label{E:MichMen}
y_i= \frac{1}{1+k/x_i}
\end{equation}
(best known as  \citet{Michaelis1913} equation in biochemistry) plotted semi logarithmically \citep{Michaelis1913} with uncertainty added, as follows
\begin{equation}\label{E:MichMenErr}
y_i= \frac{1}{1+k/x_i} + N(0,0.1)
\end{equation}
for 20 data sets with abscissa data points $\{x_i\}=\lbrace 0.01, 0.03, 1, 3, 10, 30 , 100 , 300 \rbrace$ with $ k= 5$; the results are presented as diverse symbols in Fig. \ref{fig:Figure1}B. The solid line in Fig. \ref{fig:Figure1}B obeys Eq.  (\ref{E:BoltzElect}), with $V_{1/2}=0.716864$ and $\kappa=0.430228$, both values determined with the simplex algorithm for $n = 200$ pairs of simulated data in the decimal semilogarithmic plane $\lbrace \log_{10}(x_i) , y_i \rbrace$, where Michaelis--Menten data [Eq. (\ref{E:MichMen})] become a sigmoid function as seen in Fig \ref{fig:Figure1}B. 

The right model--right data relationship in the case of Fig. \ref{fig:Figure1}A is clear; random variables from a Boltzmann process are fitted with a Boltzmann equation, and the fit is quite good. The case of Fig. \ref{fig:Figure1}B is different, a Michaelis--Menten process is not a equal to a Boltzmann process (for a further discussion see Appendix \ref{S:Yifrach}), and yet the Boltzmann model represented by the solid line seems to fit the data in Fig. \ref{fig:Figure1}B well, and predicts correctly several properties of the random Michaelis-Mented process used to generate the data, $ \displaystyle\lim_{x_i \to \infty} y = 1$ and $ \displaystyle\lim_{x \to 0} y = 0$. As mentioned before $\kappa$ is the slope of the Boltzmann function when $x \rightarrow V_{\text{\textonehalf}}$. 
The value predicted with the data in Fig. \ref{fig:Figure1}B ($\kappa=0.430228\ldots$), overestimates  the real value of $k$ used to calculate the data [Eq. (\ref{E:DecSlope})] by $\approx25$\%. since $k \approx 10^{0.716864\ldots} \simeq 5.21\ldots$. It follows that, in spite of the apparently good fit between the data and the model in Fig. \ref{fig:Figure1}B, Eq. (\ref{E:BoltzElect}) not only lacks any mechanistic meaning regarding the data in the figure, but could lead to estimate a slope value which is wrong and, if used for this purpose, to assume a wrong molecularity for the Michaelis--Menten kind of reaction.

When Eq. (\ref{E:logistic}) is plotted in semilogarithmic coordinates (for $ x>0 $), it becomes sigmoidal, and it slope is
\begin{equation}\label{E:MMSlope}
\left [\frac{dy}{d[\ln(x)]} \right ]_{x=k} = \left [x  \frac{dy}{dx}\right ]_{x=k}=\frac{\eta \;  y_{m}}{4}
\end{equation}
at the point where $x_i = k$. Then, with $\eta = 1$ and $y_{m}=1$, the following is true for the form used in Eq. (\ref{E:MichMen}) and Fig. \ref{fig:Figure1}B
\begin{equation}\label{E:DecSlope}
\left [\frac{dy}{d[\log_{10}(x)]} \right ]_{x=k} = \left [\frac{x}{\log_{10}(e)} \cdot  \frac{dy}{dx}\right ]_{x=k}=\frac{\eta \;  y_{m}}{4 \, \log_{10}(e)} \simeq 0.5756\ldots
\end{equation}

\begin{figure}
\centering
\includegraphics[width=12cm]{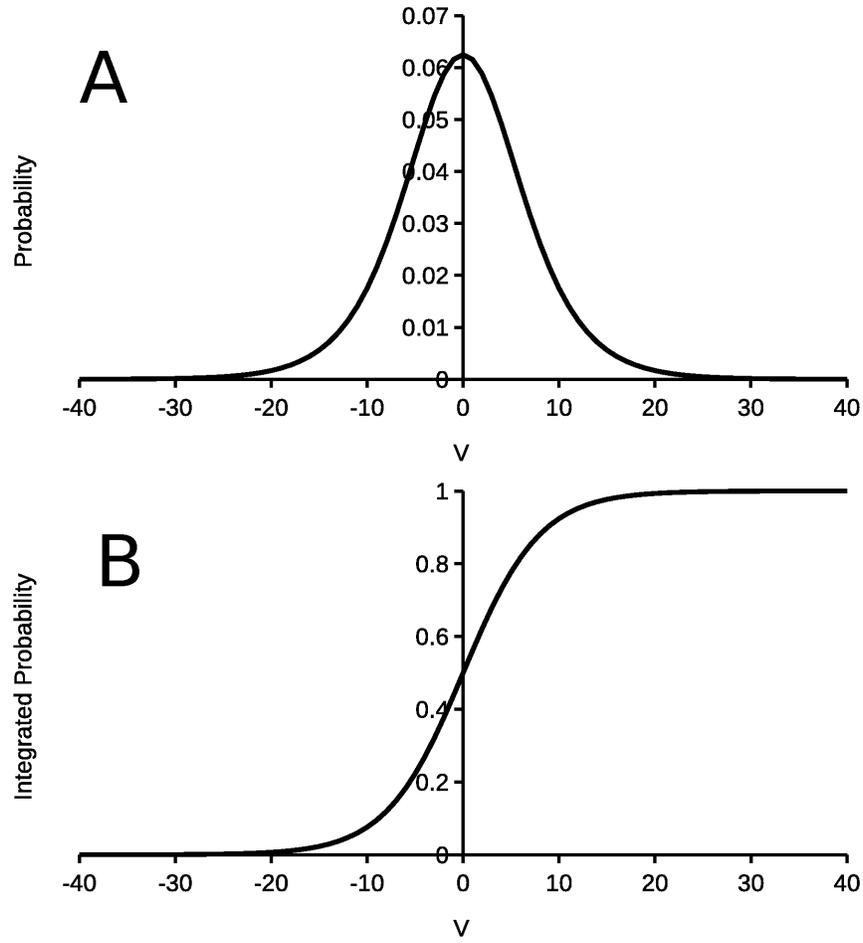}
\caption{\textbf{The Boltzmann function and its derivative probability density function (pdf)}. \textbf{Panel A: }The pdf associated with Eq. (\ref{E:BoltzElect}) as expressed by Eq. (\ref{E:Boltz_pdf}). \textbf{Panel B:} Is the probability distribution function (PDF) calculated as Eq. (\ref{E:BoltzElect}). In both cases the functions were plotted with $V_{\text{\textonehalf}}=0$ and $\kappa=4$.}
\label{fig:Figure2}
\end{figure}

  \begin{table}
  \begin{center}
  
  \topcaption{\textbf{Parameters characterizing a sample of 200,000 $w$ points generated as indicated by Eqs. (\ref{E:wTransf}) and (\ref{E:Moments}) }}\label{tab:SamplePar}
  
  \begin{supertabular}{m{4.8cm}*{3}{m{4cm}}}
  \hhline{*{4}=}
  \textbf{Parameter} & \textbf{Value} & \textbf{SEM }  & \textbf{95 \% CI}  \\
  \hhline{*{4}-}
  \textbf{Sample mean} & $-2.75364 \cdot 10^{-2}$  & $1.620 \cdot 10^{-2}$  & $(-5.912, 0.404)\cdot 10^{-2}$ \\
  \textbf{Theoretical mean}   & 0 & $1.622  \cdot 10^{-2}$ & $(-3,164, 3,164)\cdot 10^{-2}$ \\
  \textbf{Sample variance}  & 52.4558 &  \\
  \textbf{Theoretical variance}  & $52.6378\ldots$  &  &  \\
  \textbf{Sample skewness} & $-7.88 \cdot 10^{-3}$ &  &  \\
  \textbf{Theoretical skewness} & $0$  &  &  \\
  \textbf{Sample kurtosis}  & $4.205$  & &  \\
  \textbf{Theoretical kurtosis}  & $4.2$  &  &  \\
  \hhline{*{4}=} 
  \end{supertabular} 
  \end{center}
  \begin{footnotesize}
  SEM = standard error of the mean; CI = confidence interval. Theoretical parameter values are: mean and variance are moments $\omega_1$ and $\omega_2$ in Eq. (\ref{E:FourMom}); skewness is $S$ in Eq. (\ref{E:skewness}): kurtosis is $K$ in Eq. (\ref{E:kurtosis}).
  \end{footnotesize}
  \end{table}

\subsection{Comparing the theoretical Boltzmann function with values obtained in Monte Carlo simulations.}

Two hundred thousand variables distributed as $b(w) $ were generated by rearranging Eq. (\ref{E:BoltzElect}) as
\begin{equation}\label{E:Generbw}
w_i = - \kappa \, \ln \frac{1- u_i}{u_i}
\end{equation}
where $u_i$ was a random number uniformly distributed in $[0,1]$ calculated with the genrand\_res53() function of \citet{Matsumoto1998}. Some statistical properties of this set are shown in Table  \ref{tab:SamplePar}. The discrepancies between estimated sample parameters and the theoretical values in the table are very small, and most of them are related to the numerical approximations needed in computer simulation and the precision used to store the test file, set to 15 significant digits here; the discrepancies are reduced when \textbf{long double} functions are used instead of \textbf{double} functions in the C++ program, for example.

\begin{figure}
\centering
\includegraphics[width=12cm]{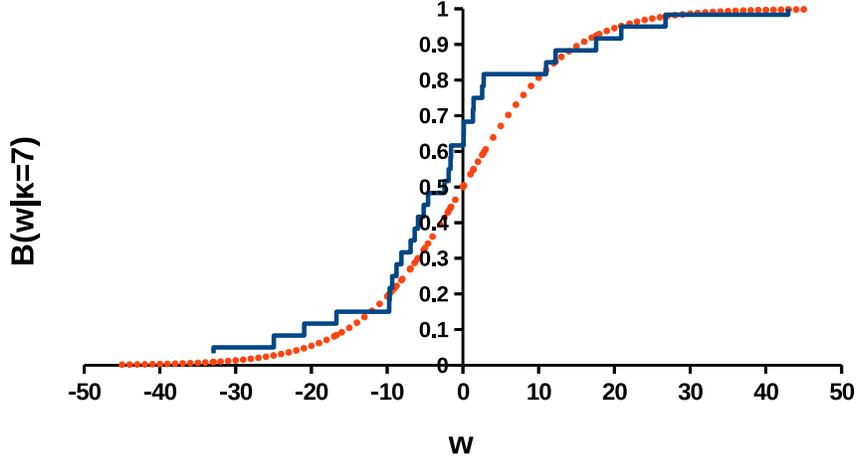}
\caption{\textbf{Boltzmann PDF (dotted line) and an empirical PDF (stepped line) determined from 30 points produced by Monte Carlo simulation using the Boltzmann PDF.} The dotted curve was generated as $ B(w|\kappa)= 1/(1+e^{-w/\kappa}) $. The stepped line is the empirical PDF calculated via Monte Carlo simulation for a sample of 30 points generated as $w_i= \kappa \ln\left[ (1-u_i)/u_i\right]  $ [Eq. (\ref{E:Generbw}) in te text of this work], where $u_i=U[0,1] $ a uniform random variable in the closed interval $ [0,1] $. In this figure $ \kappa=7 $. Arrow indicates maximum distance between empirical and Boltzmann PDFs. See the text for other details.}
\label{fig:Figure3}
\end{figure}

Table \ref{tab:SamplePar}, does not stress enough that Boltzmann processes \textit{per se} are highly stochastic independently of the inherent uncertainty of any data collecting empirical procedure; the variability of $ \{w_i\} $ depends only on $ \kappa  $ as indicated by the 2\textsuperscript{nd} moment ($ \omega_2 $) in equation set (\ref{E:FourMom}). Figure \ref{fig:Figure3} simulates an experiment where 30 values of $\{w_i\}$ obeying Eq. (\ref{E:BoltzElect}) were recorded and an empirical PDF (\textit{EDF}) was constructed (stepped line) for the set $ \{w_i\}_{i=1,2,\ldots,30} $, in both cases $ \kappa=7 $. The EDF is a step function that jumps up by $1/n$ at each of the $n$ data points [Eq. (\ref{E:EDF})]. The EDF estimates the PDF underlying the points in the sample and converges with probability 1 according to the Glivenko-Cantelli \citep{Glivenko1933, Cantelli1933} theorem. As indicated in the figure, the EDF for 30 points which has no measurement uncertainty, deviates significantly from the theoretical PDF, \emph{due only to the stochasticity inherent to the Boltzmann PDF}. The parameters for the simulations in Figure \ref{fig:Figure3} were chosen to mimic $ \kappa $ values commonly observed in electrophysiology \citep[pg 501, Eq. 1]{Hodgkin1952d}, but could be any other set of values. The maximum difference between the 2 curves in Figure \ref{fig:Figure3} is 0.203 (Arrow in Figure \ref{fig:Figure3}), the Smirnoff test \citep{Smirnov1939, Smirnov1948} in used to compare the two curves the probability that the differences between the two PDFs in the figure stems from chance will be estimated as $ P < 10^{-6} $, which leads to a statistical error of type I, the incorrect rejection of a true null hypothesis. Under experimental conditions, usually 15 to 20 points are recorded in a single experiment which is replicated in some 3 to 5 subjects (or preparations), under conditions where preparation deterioration, recording noise or instrumental limitations add uncertainty to the variable studied. The situation is further made prone to errors by rigidly setting the null hypothesis rejection \textquotedblleft{}threshold\textquotedblright{} at $ P \leq 0.05 $ ignoring that in many instances this \textquotedblleft{}threshold\textquotedblright{} depends on the stochastic variable under consideration \citep{Bonferroni1936,Ioannidis2005,Colquhoun2014}.

\section{Discussion.}

\textit{Determinism} is a metaphysical philosophical position stating that for everything that happens there are conditions such that, given those conditions, nothing else could happen. Many mathematical models of physical systems are deterministic. Mathematical models which are not deterministic because they involve randomness are called \textit{stochastic}. Even systems that involve no randomness involve \textit{uncertainty}, for reasons such as: 
\begin{enumerate}
\item Limitations of the observation instrument make the measurements fuzzy. Optical instruments are a very intuitive example of fuzziness. Actually, measuring instrument introduces fuzziness, usually called \textit{uncertainty}, which is dependent on the instrument\textquoteright{}s \textit{limit of resolution}.
\item Members of a population are not all equal. Individual variability and biological diversity are essential to life.
\item Observing reality with a scope (aim or purpose) modifies the object observed \citep{Sassoli2013}. This is specially relevant to quantum physics, but applies to any measurement (draining current, compressing with a caliper, heating, etc.) to, hopefully, a minor extent.
\item The object measured changes more or less cyclically in time. The height of the Mont Blanc peak (like most other mountains) is a well known case \citep{Gilluly1949,Evans2015}.
\item The dimension of the observed object depends on the scale of measurement (fractals) \citep{Mandelbrot1983}. Coastlines, river beds or lengths, national borders, are classical examples.
\item  The system studied is deterministic but unpredictable (chaotic), due to strong nonlinearity or sensitivity to initial conditions, such as weather, population growth, seismic activity and many more \citep{Lorenz1963, May1976, Feigenbaum1978, Feigenbaum1980a, Feigenbaum1980b, Gonzalez2000, Sevcik1976a}.
\item The variables measured are any of a variety of mathematical inequalities asserting a fundamental limit to the precision with which certain pairs of physical properties of a particle, can be known simultaneously \citep{Heisenberg1927}.
\end{enumerate}

Equations of the form of Eq. (\ref{E:BoltzElect}) have become extremely popular to describe diverse biological situations \citep{Hodgkin1952d, Rouzaire1991, Fogolari2002, Koegel2003, Habela2007, Schuster2008, Zhou2008, Cambien2008, Dubois2009}. The purpose of this paper is to explore the plasticity of the Boltzmann function to fit data, some aspects of the optimization procedure to fit the function to data and on how to use this plastic function to differentiate the effect of changing system conditions, also called: \textit{treatment}. 

Due to the reasons discussed in connection with Eq.(\ref{E:HessMin}), neither linear transformation nor nonlinear regressions are useful to estimate the uncertainty of unknown, linearly dependent, parameters $\{V_{\text{\textonehalf}},\kappa\}$ of Boltzmann  functions, fitted to data; their variances due to intra-experimental uncertainty cannot be separated and cannot be measured independently. This shortcoming is more serious given that, as shown in Section \ref{S:Plasticity} and Figure \ref{fig:Figure1}, Boltzmann functions may be fitted to data for which the function has no mechanistic model value. In this regard, it must be stated that rectangular hyperbolas (which was chosen arbitrarily as just an example), such as the Michaelis--Menten equation \citep{Michaelis1913} or its extention, the Hill equation \citep{Hill1910b} are not related to de Boltzmann function in spite of some claims on the contrary \citep{Yifrach2004}, as it is discussed in more details in Appendix \ref{S:Yifrach}.

 It has been said that \textquotedblleft{}many physicists have little knowledge of statistics.\textquotedblright $\ldots$ \textquotedblright{}this mostly arises because there is little need for statistics in physics\textquotedblright{} \citep{Chow2009}. The main difference between physics and biology is that physics methods are applied to measure phenomena that are not too uncertain; most physical constants are known precisely to well beyond 5 decimal places. But as high resolution modern physics reaches its resolution limits (such as the most significant finding in recent times, the existence of the Higgs boson) it depends critically on separating a small signal from its surrounding uncertainty by statistical means \citep[Figure 3]{CERN2013}. 
 
 Variability and diversity are fundamental for preserving life; consequently, biological phenomena may be known only with significant uncertainty, and biological parameters cannot be usually known with more that 5\% accuracy. The problem is certainly worse in biology and medicine, where experimenters may have little mathematical background, or even may  be skeptic about the value of mathematics, statistics and mathematical modeling. Quoting \citet[1.1. How to avoid making a fool of yourself. The role of statistics]{Colquhoun1971}:
\begin{quote}
It is widely held by non-statisticians, like the author, that if you do good experiments statistics are not necessary. They are quite right. At least they are right as long as one makes an exception of the important branch of statistics that deals with processes that are \textit{inherently} statistical in nature, so called \textquoteleft{}\textit{stochastic}\textquoteright{} processes\textquotedblright{} $\ldots$. The snag of course is that doing good experiments is difficult. Most people need all the help they can get to prevent them making a fool of themselves by claiming that their favourite theory is sustained by observations that do nothing of the sort.
\end{quote}
 In biology, a set of $m$ experiments is performed,  then a Boltzmann or other function is fitted to each experiment, and the values of the unknown model parameters ($V_{\text{\textonehalf}}$ and $\kappa$, for example) obtained under different experimental conditions, are statistically compared without taking into account the intra experimental uncertainty stemming from the dispersion of data about the function fitted. Usually the experimenters are unaware of that such practice is prone to result in statistical type I error, which occur when the null hypothesis is true, but is rejected; this is asserting something that is absent, a false hit. A type I error may be compared with a so-called false positive (a result that indicates that a given condition is present when it actually is not present) \citep{Wilks1962,Colquhoun2014,Loiselle2015}. Consequently, it is wise to distrust small or weakly statistically significant differences between Boltzmann function parameters fitted to data under different experimental conditions.
 
The handicap resulting from being unable to estimate intra experimental uncertainty of unknown model parameters under, say, two experimental conditions, is most annoying when the question to answer is: Is $\theta_{i,1}$ different from $\theta_{i,2}$?. This is so since it the uncertainty of $\theta_{1,j}$ cannot be separated from any other $\theta_{i\neq1,j}$. Yet under the condition of \textit{paired comparisons}, when \textit{the same system }is studied under two experimental conditions which may be described with equations like Eq. (\ref{E:BoltzElect}), say  $B(V|V_{{\text{\textonehalf}}, 1},\kappa_1)$ and $B(V|V_{{\text{\textonehalf}}, 2},\kappa_2)$, the question: Is $B(V|V_{{\text{\textonehalf}}, 1},\kappa_1)$ different from $B(V|V_{{\text{\textonehalf}}, 2},\kappa_2)$?, may be answered more accurately (with less likelihood of statistical errors of type I or II \citep{Wilks1962}). This is so since Eq. (\ref{E:BoltzElect}) is a PDF, and data described by the equation under different, \textit{but paired}, experimental conditions may be used to build EDFs which may be compared (see Section \ref{S:pdfPDF}), no matter to which PDF they correspond (Section \ref{S:CompBoltz} of this communication), using Kolmogorov-Smirnov statistics \citep[see Ch. 6 for practical aspects]{Conover1999}. If Kolmogorov-Smirnov statistics indicates that treatment modifies the system studied, but the uncertainty about which parameter(s) is(are) changing will not be dispelled. A nonparametric statistical comparison between individual model variables, combined with the Kolmogorov-Smirnov test, will reduce uncertainty to assess treatment efficacy more precisely than any of the two approaches \textit{per se} \citep{Ariens1964}.

\section{Appendixes}

\begin{appendices}
	
\section{On square hiperbolas and the Boltzmann function.}\label{S:Yifrach}

It is possible to deduce the Michaelis--Menten equation [Eq. \ref{E:MichMen}] in a more general form \citep[Pg. 471, Eq. (VIII-9)]{Segel1975}, this form is sometimes called \textit{general logistic function} \citep[Pg. 32]{Kenakin1997} or the Hill \textit{equation} \citep{Hill1910b, Hill1913,Wyman1951}
\begin{equation}\label{E:logistic}
y=\frac{y_{m}  x^{\eta}}{x^{\eta}+ k^{\eta}} = \frac{y_{m}}{1+\left (k/x\right )^\eta}= \frac{y_{m}}{1+q/x^{\eta}}
\end{equation}
where $\eta$ is interpreted as the number of drug molecules interacting with a receptor (the \textit{molecularity} of the reaction), and $y_{max}$ is the maximum drug effect or catalytic rate. Equation (\ref{E:logistic}) may be linearized as
\begin{equation}\label{E:HillTr}
\log{\dfrac{y_m-y}{y}}=\log{q}-\eta \log{x} \; \equiv \; \varUpsilon_1=A_1+B_1 X_1
\end{equation}
a linear transformation which is a log-log plot of an equation in \citet[3\textsuperscript{rd} equation in page 447]{Hill1913} and is known as the Hill \textit{plot} \citep{Resnick1997,Drees2000,Lehninger2013} (\textquotedblleft{}$ \equiv $\textquotedblright{} indicates equivalence between the equations) which is of great importance in cooperative binding analysis \citep{Bohr1904, Stefan2013}. An unrelated Hill \textit{estimator} is also used in statistics \citep{Vollmer2001}. From the Hill equation $\eta$ is also called the Hill \textit{coefficient}. 

It  was suggested by \citet{Yifrach2004} that a logarithmic transform of a Boltzmann function  \citep[Eq. (7)]{Almers1978}\citep[Eq. (1)]{Yifrach2004}
\begin{equation}\label{E:Yifrach1}
K(V) = \frac{O}{C}=K e^{\frac{Z_TFV}{RT}}
\end{equation}
is equivalent to the Hill \textit{plot}. In Eq (\ref{E:Yifrach1}) the equilibrium between the open ($ O $) an closed ($ C $) states of an ionic channel is voltage ($ V $) dependent. Other parameters in Eq. (\ref{E:Yifrach1}) are: $ K $ is the chemical equilibrium constant for an ion channel gating in the absence of voltage (at 0 mV); $ Z_T $ is the total gating charge of a channel that moves across the membrane electrical field upon depolarization; $ F $ is the Faraday constant; $ R $ is the gas constant and $T$ is absolute temperature. As shown by \citet[Eq. (2)]{Yifrach2004}, equation (\ref{E:Yifrach1}) may be transformed into a linear form such as
\begin{equation}\label{E:Yifrach2}
	\log \dfrac{P}{1-P}=\log K + n_H \dfrac{Z_U F}{RT}V \; \equiv \; \log \dfrac{1-P}{P}=\log Q - n_H \dfrac{-Z_U F}{RT}V  \; \equiv \; \varUpsilon_2 =A_2 + B_2  X_2 \text{,}
\end{equation}
where $ P $ is the probability of the channel being open [$ P=O/(O+C) $], $ n_H $ is the number of channel subunits, and $ Z_U $ is the unitary gating charge associated with the subunit [$ Z_T=n_H Z_U $]. \citet[Eq. 839]{Yifrach2004} uses a sophism to state
\begin{equation}\label{E:Yifrach3}
\log \dfrac{\overline{Y}}{1-\overline{Y}}=\log K + n_H \log(S) \; \equiv \;\log \dfrac{1-\overline{Y}}{\overline{Y}}=\log Q - n_H \log(\text{\textit{\ss{}}})
\end{equation}
where (quoting \citet{Yifrach2004})
\begin{quote}
$ \overline{Y} $ ; the fractional binding saturation function, is the fraction of sites occupied with the substrate ($ S $), $ n_H $ is the Hill coefficient, and $ K $ is the apparent binding constant of the substrate to the enzyme.
\end{quote}  
Obviously $ S= \text{antilog}\left (\frac{Z_U F}{RT}V \right )$ but its use as in Eq. (\ref{E:Yifrach3}) is a tautology (by definition a variable is the inverse of its inverse, when the inverse exists), a change of variables that turns a \textit{semilog} plot into a \textit{log-log} plot. \citet{Yifrach2004} reasoning in fact proves the opposite of its intention: Eqs. (\ref{E:logistic}) and (\ref{E:Yifrach1}) are \textit{different} since to make them \textit{lookalike} you have to transform them into \textit{different} spaces. Thus in spite of a \textit{lookalike} condition between Eqs. (\ref{E:Yifrach1}) and (\ref{E:Yifrach2}), reinforced by the sophistical Eq. (\ref{E:Yifrach3}), the equations are different, not isomorphic. Eq. (\ref{E:HillTr}) is a \textit{log-log} transformation of Eq. (\ref{E:logistic}) where $ \log[(y_{max} - y)/y] $ is plotted \textit{versus} $ \log(x) $, whereas Eq.(\ref{E:Yifrach2}), is a \textit{semilog} plot of $ \log [(1-P)/P] $ \textit{versus} $ V $. Thus, neither in its original form [Eq. (\ref{E:logistic})] nor in its transformed [Eq. (\ref{E:HillTr})] form, the logistic (Hill) Eq. (\ref{E:logistic}) corresponds to Boltzmann Eq. (\ref{E:Yifrach1}) and \citet{Yifrach2004} choice of calling $ n_H $ \textquotedblleft{}a Hill coefficient\textquotedblright{} is forced and misleading. 

\section{Mathematical appendix.}

\subsection{How does the Bolzmann function compare with a probability distribution function?}\label{S:pdfPDF}

The Boltzmann function in Eq. (\ref{E:MaxwBoltz})  through (\ref{E:Gating1}) form, tells how likely it is that $N_j$ particles are in a given state out of $N$ possible states, not the probability that a given particle is in such state. It is definitely not a PDF. A condition that any continuous PDF, say $f(x)$, must fulfill is $\int_{-\infty}^{\infty} f(x) dx\,=\,1$  which is not the case of the Boltzmann PDF.

 For any continuous pdf, such as $f(x)$,  $F(w)=\int_{-\infty}^{w} f(x) \; dx $ exists and is a random variable $U[0,1]$ uniformly distributed in the interval [0,1], the so called \textit{standard uniform distribution} $U[0,1]$, sometimes also referred to as \textit{rectangular distribution} $R(\mu, \xi) = R(\text{\textonehalf}, 1)$ \citep[pg, 155]{Wilks1962} since it has an expectation $\mu = \text{\textonehalf}$ and and any value in the closed interval $[0,1]$ occurs with a probability $\xi= 1$. . This is demonstrated by the following theorem \citep[pg. 156, Theo. 7.1.1]{Wilks1962}. 
 
\begin{theorem}[\citet{Wilks1962}, pg. 156]\label{T:PDF_Unif}
If $x$ is a random variable having a PDF $F(x)$ then the random variable $y = F(x)$ has the  rectangular distribution $R(\text{\textonehalf}, 1)$.
\begin{proof}
This follows at once from the fact that the PDF of $y$ is 
\begin{equation}
G(y)= P \left [ F(x) \leqslant y \right ] =  \begin{cases} 1, &  y > 1   \\  y, & 0 < y \leqslant 1 \\  0, & y \leqslant 0 \end{cases}
\end{equation}
which is the pdf of the rectangular distribution $R(\text{\textonehalf}, 1)$.
\end{proof}
\end{theorem}

Theorem (\ref{T:PDF_Unif}) is obviously true for the Boltzmann function such as Eqs. (\ref{E:MaxwBoltz})  through (\ref{E:Gating1}). Thus the Boltzmann function ($B(V|V_{\text{\textonehalf}},\kappa)$) is a PDF, of the pdf
\begin{equation}\label{E:Boltz_pdf}
\frac{dB(V|V_{\text{\textonehalf}},\kappa)}{dV} = b(V|V_{\text{\textonehalf}},\kappa))=\frac{e^{-(V-V_{\text{\textonehalf}}) / \kappa}}{\kappa \left [1+e^{-(V-V_{\text{\textonehalf}})/\kappa} \right ]^2}=\frac{1}{2 \kappa \left[1 + \cosh\left (\frac{V-V_{\text{\textonehalf}}}{\kappa}  \right ) \right]}\; .
\end{equation}
subject to necessary condition that $(\kappa \neq 0) \in \mathbb{R}$, where $\mathbb{R}$ is the set of real numbers, then $\int_{-\infty}^{ \infty} b(V|V_{\text{\textonehalf}},\kappa)) \, dV = 1$ and Eq. (\ref{E:Boltz_pdf}) hold.

\subsection{Boltzmann PDF central moments.}\label{sec:BoltzMom}
Two PDFs are equal if they have the same  moment generating function (\textit{MGF}) \citep{Wilks1962}. The MGF  of the Boltzmann PDF may be obtained as follows: 
\begin{equation}
\because \quad b(w|V_{\text{\textonehalf}},\kappa) = b(V -V_{\text{\textonehalf}}|V_{\text{\textonehalf}},\kappa) \label{E:wTransf} 
\end{equation}
and since the MGF is generally defined as $M_w(t) = \mathbb{E}\left (e^{tw}\right )=\int \limits_{ -\infty}^{\infty}e^{tw}dx$ then 
\begin{equation}
M_w(t) =1+ \sum \limits_{i=1}^{\infty} t^i \dfrac{\mathbb{E}(w^{i})}{i!}=1+ \sum \limits_{i=1}^{\infty} t^i \dfrac{\omega_{i}}{i!}
\end{equation}
\begin{equation}
\therefore \quad \omega_k =  \mathbb{E} \left( w^k \right )  \label{E:Moments}
\end{equation}
is the\textit{ k\textsuperscript{it}} \textit{central moment} of $w$. Then, if $(\kappa ^{-1} >0) \in \mathbb{R} $ the first four central moments, $\omega_k$, of $b(w|V_{\text{\textonehalf}},\kappa)$ are 
\begin{align}\label{E:FourMom}
\omega_1 &= \mu= 0 &
\omega_ 2 &= \sigma^2 =  \frac{\pi^2 \, \kappa^2}{3} \\ 
\omega_3 &= 0 &
\omega_4 &= \frac{7 \, \pi^4 \, \kappa^4}{15} \nonumber
\end{align}
As may be appreciated in Fig. \ref{fig:Figure2}A the Boltzmann PDF described by Eq. (\ref{E:Boltz_pdf}) is symmetric about $w=0$.  The  skewness \citep[see pg. 265 for details on  skewness]{Wilks1962} of $b(w)$ is 
\begin{equation}\label{E:skewness}
 S=\frac{\omega_3}{\omega_2^{3/2}}=0\text{.}
 \end{equation}
  Although the plot in Fig. \ref{fig:Figure2}A evokes the Gaussian bell, the distributions are different, $b(x)$ has a kurtosis \citep[see pg. 265 for details on kurtosis]{Wilks1962} exactly equal to
  \begin{equation}\label{E:kurtosis}
  K=\frac{\omega_4}{\omega_2^{2}}=4.2\text{,}
  \end{equation}
  higher than the Gauss\textquoteright{} pdf kurtosis which is is exactly 3.
  
\subsection{Boltzmann pdf Hessian and variance.}

The \textit{Hessian} matrix of a function $f(x|\theta_{i})_{i=1,\ldots, n}$  is a matrix of second partial derivatives of the form
\begin{equation}\label{E:GenHessian}
Hf \left(x | \theta_i \right) \,=\, \left[ 
\begin{matrix} 
\dfrac{\partial^2 f(x|\theta_i)}{\partial\theta_1^2}&\dfrac{\partial^2 f(x|\theta_i)}{\partial\theta_1\partial\theta_2}&\dots&\dfrac{\partial^2 f(x|\theta_i)}{\partial\theta_1\partial\theta_n}\\
\dfrac{\partial^2 f(x|\theta_i)}{\partial\theta_2\partial\theta_1}&\dfrac{\partial^2 f(x|\theta_i)}{\partial\theta_2^2}&\dots&\dfrac{\partial^2 f(x|\theta_i)}{\partial\theta_2\partial\theta_n}\\
\vdots&\vdots&\ddots&\vdots\\
\dfrac{\partial^2 f(x|\theta_i)}{\partial\theta_n\partial\theta_1}&\dfrac{\partial^2 f(x|\theta_i)}{\partial\theta_n\partial\theta_2}&\dots& \dfrac{\partial^2 f(x|\theta_i)}{\partial\theta_n^2}
\end{matrix} \right]
\end{equation}
If $\{\theta_i\}$ are all linearly independent, then $Hf \left(x | \theta_i \right)$ is the diagonal matrix:
\begin{equation}\label{E:HessianInd}
Hf \left(x | \theta_i \right)\,=\,
\left[ 
\begin{matrix} 
\dfrac{\partial^2 f(x|\theta_i)}{\partial\theta_1^2}&0&\dots&0\\
0&\dfrac{\partial^2 f(x|\theta_i)}{\partial\theta_2^2}&\dots&0\\
\vdots&\vdots&\ddots&\vdots\\
0&0&\dots& \dfrac{\partial^2 f(x|\theta_i)}{\partial\theta_n^2}
\end{matrix}  \right]
\end{equation}

Where $\{ \gamma_i  \}  = \left \{   \dfrac{\partial^2 f(x|\theta_i)}{\partial\theta_i^2} \right \}_{i=1,\ldots,n}$
 are the matrix \textit{eigenvalues}.

The Hessian matrix, of Eq. (\ref{E:BoltzElect}) is the non diagonal matrix:
\begin{equation}\label{E:Hessian}
HB\left (w|V_{\text{\textonehalf}},\kappa \right )\,=\, \dfrac{\text{sech}^2\left(\xi\right)}{4 \kappa ^4}\left [
\begin{matrix}
\dfrac{8\kappa^2 \sinh ^4\left(\xi\right) \text{csch}^3\left(2\xi\right)}{\text{sech}^2\left(\xi\right)}&\kappa\left[\kappa -w \tanh \left(\xi\right)\right]  \\ \kappa\left[\kappa -w\tanh \left(\xi\right)\right]&w\left[2 \kappa w \tanh \left(\xi\right)\right] 
\end{matrix} \right ]
\end{equation}
where $\xi=\dfrac{V-V_{\text{\textonehalf}}}{2\kappa}=\dfrac{w}{2\kappa}$;  $HB(w|V_{\text{\textonehalf}},\kappa ) $ is symmetric and non singular, which indicates that $\{V_{\text{\textonehalf}},\kappa\}$ are not linearly independent. Also
\begin{equation}\label{E:HessMin}
HB(w=0|V_{\text{\textonehalf}},\kappa )=\dfrac{1}{4}\left [
\begin{matrix} 
0&1 \\ 
1&0 
\end{matrix}\right ] =\dfrac{1}{4}\left [
\begin{matrix} 
-1&0 \\ 
0&1 
\end{matrix}\right ]
\end{equation}
which evaluating the matrix as a determinant (called the \textit{discriminant}), has a value of -\textonequarter{} with eigenvalues \{-1\,,\,1\}. It is a negative definite matrix; meaning that no mater what is the value of $\kappa$, the surface containing all the solutions of $B(w|V_{\text{\textonehalf}},\kappa )$ has a maximum at $w=0$ which is a \textit{critical point} of $B\left (w|V_{\text{\textonehalf}},\kappa \right )$.

\subsection{Kolmogorov distribution functions.}\label{S:CompBoltz}
If $\lbrace x_i \rbrace $ is a set of $n$ random variables ordered so that $\lbrace x_1 \leqslant x_2 \leqslant \ldots \leqslant x_n \rbrace $. Then $F_{n}(x)$, called by \citet{Kolmogorov1933} an EDF, has the following properties:
\begin{equation}\label{E:EDF}
F_{n}(x) = \begin{cases} 0,   & x < x_1 \\
\dfrac{i}{n},   &  \{x_i \leqslant x \leqslant x_{i+1}\}_{ i = 1,  2, \ldots ,  n-1}\\
1,   &  x_n \leqslant x  \end{cases}
\end{equation}
then to measure how close $F(x)$, a distribution function, is from $ F_{n}(x) $ \citet{Kolmogorov1933} defined the supremum
\begin{equation}\label{E:DFnF}
D\;=\;\underset{x}{\sup}\left|{F_{n}(x)\;-\;F(x)}\right|
\end{equation}
where the bars indicates absolute values, \textit{x} is the point where the supremum occurs. By the Glivenko–Cantelli theorem \citep{Cantelli1933, Glivenko1933}, if the sample comes from distribution $F(x)$, then $D$ converges to 0 almost surely. \citet{Kolmogorov1933} strengthened this result, by effectively providing the rate of this convergence. The following two theorems are described inspired on Theo. (1) and (2) in \citet{Feller1948}).

\begin{theorem}[\citet{Kolmogorov1933}]\label{T:Kolmogorov}

For every fixed $\lambda >0$ if  $Pr(D < \lambda/ \sqrt{n})$ denotes the probability that $D < \lambda/ \sqrt{n}$ then
\begin{equation}
\lim_{n \to \infty} Pr \left (D <\frac{ \lambda}{ \sqrt{n}}\right ) = L(\lambda)
\end{equation}
where $L(\lambda)$ is the probability distribution function (pdf), which for $\lambda > 0$, is given by either of the following equivalent relations
\begin{equation}
L(\lambda)=1 - \sum_{k= -\infty}^{\infty} (-1)^k e^{-2k^2\lambda^2} =  \frac{\sqrt{2\pi}}{\lambda}\sum_{k=1}^{\infty}e^{-\frac{(2k-1)^2 \pi^2}{8\lambda^2}}
\end{equation}
For $\lambda \leq 0$, $L(\lambda)=0$.
\end{theorem}

The \citet{Kolmogorov1933} proof of Theo. (\ref*{T:Kolmogorov}) demands that
\begin{equation}\label{E:CumulProb}
\Phi(y)=\int \limits_{-\infty}^{y}F(x)~dx
\end{equation}
is a uniform random variable such as $U[0,1]$. Equation (\ref{E:CumulProb}) characterizes the PDF of any continuous pdf.

Another theorem, due to \citet{Smirnov1939} states:

\begin{theorem}[\citet{Smirnov1939}]\label{T:Smirnov}
Let $\lbrace x_i \rbrace_j $ be a sets of $n_j$ mutually independent random variables ordered so that $\lbrace x_1 \leq x_2 \leq \ldots \leq x_{n_j} \rbrace_j $ for which $F_{j, n_j}(x)$ may be defined. Then if
\begin{equation}
D_{n_1,n_2}\;=\;\underset{x}{\sup}\left|{F_{n_1}(x)\;-\;F_{n_2}(x)}\right| 
\end{equation}
define
\begin{equation}
\nu= \frac{n_1 n_2}{n_1 + n_2}
\end{equation}
and suppose that
\begin{equation}
\lim_{n_1 \to \infty, n_2 \to \infty} \frac{n_1}{n_2} = a
\end{equation}
where $a$ is a constant. Then for every fixed $\lambda > 0 $
\begin{equation}
Pr(D < \lambda/ \sqrt{\nu}) =  1 - \sum_{k= -\infty}^{\infty} (-1)^k e^{-2k^2\lambda^2} =  \frac{\sqrt{2\pi}}{\lambda}\sum_{k=1}^{\infty}e^{-\frac{(2k-1)^2 \pi^2}{8\lambda^2}} .
\end{equation}
\end{theorem}

\noindent{Theorem (\ref{T:Smirnov}) is a generalization of Theo. (\ref{T:Kolmogorov}) to compare two empirical distribution functions based on \citet{Kolmogorov1933} statistics and is the base of the so called Kolmogorov-Smirnov test.}

\subsection{The Kolmogorov-Smirnov statistics in connection with the Boltzmann function.}
If $n_j$ points are drawn to produce a sample $ \lbrace \beta_{j}(V_i)\rbrace_{i=1, \ldots, n_j} $ from a process described by
\begin{equation}\label{E:ExpBoltz}
\beta_{j}(V_i) = \frac{1}{1+e^{-(V_i -V_{1/2, j}) \;/\; \kappa_j}}
\end{equation}
are ordered as $\lbrace \beta_{j}(V_1)  \leq \beta_{j}(V_2) \leq \ldots \leq \beta_{j}(V_i) \leq \ldots  \leq \beta_{j}(V_{n_j}) \rbrace_j$ where all $\beta_{j}(V_i)$ comply with the conditions in  equation set (\ref{E:EDF}) and all have the form
\begin{equation}  \label{E:RanVar}
\beta_{j}(V_i) = B_j(V_i) + \epsilon_i
\end{equation}
where $\epsilon_i$ is a random variable distributed as some PDF $h(x)$ with expectation $\int_{-\infty}^{ \infty}x\; h(x)\; dx = 0 $ and variance $ 0 < \int_{-\infty}^{\infty} x^2\; h(x)\; dx < \infty $. The subindex $ j $ introduced in equation (\ref{E:ExpBoltz}) allows for different experimental Boltzmann systems, or for the same Boltzmann system under different experimental condition often called \textit{treatments}.

Then, for two experimental samples we may define the set
\begin{equation}\label{E:DeltamInf}
\lbrace \Delta \beta_{1,2}(V)\rbrace \; = \;\left \lbrace \left |{\frac{1}{1+e^{-(V-V_{1/2,1}) \;/\; \kappa_1}} \;-\; \frac{1}{1+e^{-(V-V'_{1/2,2}) \;/ \;\kappa_2}}}\right |\right \rbrace
\end{equation}
which has a \textit{supremum},
\begin{equation}\label{E:D}
D\;=\;\underset{V}{\sup} \left \lbrace \Delta \beta_{1,2}(V)  \right \rbrace.
\end{equation}
The supremum of a subset $\{S\}$ of a totally or partially ordered set $\{T\}$ is the least element of $\{T\}$ that is greater than or equal to all elements of $\{S\}$. If the supremum exists, it is unique meaning that there will be only one supremum. The supremum expressed by Eq. (\ref{E:D}), may easily be evaluated numerically.  In practice Eqs. (\ref{E:DeltamInf}) and (\ref{E:D}) are defined only for $V \in \{ V_i\}_{i=1, \ldots, n}$, and thus, un practice, the supremum is
\begin{equation}\label{E:D_real}
D_{n,n}\;=\;\underset{V}{\sup} \left \lbrace \Delta \beta_{1,2}(V_i)  \right \rbrace.
\end{equation}
According to Kolmogorov-Smirnov statistics if $P(D_{\nu })  \leqslant  \alpha $, with  $\nu =n_1 n_2/(n_1+n_2)$, the null hypothesis may be rejected with an $\alpha $ confidence level. $P(D_{\nu })$ may be calculated with the algorithm of \citet{Marsaglia2003}, is tabulated by \citet{Conover1980} and may also be expressed as $P\left (D_{n_1,n_2} \geqslant  \frac{c}{n_1 n_2} \right  )\leqslant  \alpha $ where $c$ is tabulated by \citet[pg. 122 and Table 55]{Pearson1972}.

\end{appendices}

\section*{Acknowledgments.}

This manuscript was written in \LaTeX{ }using \textit{{\TeX}studio} for Linux (Also available for Apple OS X and MS Windows, http://www.texstudio.org), an open source free  \TeX{ }editor.

\section*{References.}


\end{document}